\documentclass[10pt, a4paper, twoside, final, onecolumn]{icton}

\IEEEoverridecommandlockouts
\usepackage{amsmath}
\usepackage{amssymb}
\usepackage{epsfig}
\usepackage{cite}
\usepackage{array}
\usepackage{placeins}
\usepackage{url}
\usepackage{enumerate}
\usepackage[centerlast]{caption}

\newcommand{\abs}[1]{\ensuremath{\lvert #1 \rvert}}

\begin{document}

\title{Beating the Standard Quantum Limit for Binary Constellations\\ in the Presence of Phase Noise}

% author names and affiliations
% use a multiple column layout for up to three different
% affiliation
\author{ \authorblockN{{\bfseries L. Kunz$^{1}$, M. T. DiMario$^{2}$, F. E. Becerra$^{2}$, and K. Banaszek$^{1}$, \textit{Senior Member, IEEE}}}\\
\authorblockA{\itshape $^1$ Centre for Quantum Optical Technologies and Faculty of Physics, University of Warsaw, Banacha~2c, 02-097 Warszawa, Poland\\
$^2$ Center for Quantum Information and Control, Department of Physics and Astronomy,\\
University of New Mexico, Albuquerque, NM 87131-0001, USA\\
e-mail: k.banaszek@cent.uw.edu.pl}\\
}

\maketitle
\thispagestyle{empty}

\pagestyle{empty}

% paper title

% no keywords

% For peer review papers, you can put extra information on the cover
% page as needed:
% \begin{center} \bfseries EDICS Category: 3-BBND \end{center}
%
% for peerreview papers, inserts a page break and creates the second title.
% Will be ignored for other modes.
%\IEEEpeerreviewmaketitle

\begin{abstract}

\noindent
Unconventional receivers enable reduction of error rates in optical communication systems below the standard quantum limit (SQL) by implementing discrimination strategies for constellation symbols that go beyond the canonical measurement of information-carrying quantities such as the intensity or quadratures of the electromagnetic field. An example of such a strategy is presented here for average-power constrained binary constellations propagating through a phase noise channel. The receiver, implementing a coherent displacement in the complex amplitude plane followed by photon number resolved detection, can be viewed as an interpolation between direct detection and homodyne detection.
\end{abstract}

\begin {keywords}
optical communication, constellation diagram, phase noise, optical receivers, coherent detection, photon counting.
\end{keywords}

\bigskip

\section{INTRODUCTION}
\noindent
In optical communication systems, information can be carried by the intensity, the phase, or generally by the complex amplitude of the electromagnetic field. Detection of any of these quantities, which conventionally serves as the basis to determine the transmitted symbol, is ultimately limited by the shot noise. Error rates resulting from the standard measurement of the information-carrying quantity performed at the shot-noise level are usually referred to as the {\em standard quantum limit} (SQL).

Quantum mechanics offers a more general perspective on the task of identifying the transmitted symbols. In quantum theory, individual symbols are associated with respective quantum states of the electromagnetic field represented mathematically by so-called density operators. Quantum fluctuations of the electromagnetic field imply that these states are not fully distinguishable in the quantum mechanical sense, i.e.\ there does not exist, even in principle, a measuring apparatus that would discriminate between them with a zero error rate. However, quantum mechanical analysis of the discrimination problem reveals that in many scenarios reduction of error rates below the SQL should be nevertheless possible \cite{Helstrom1976}.

The purpose of this contribution is to discuss a strategy to beat the SQL in a scenario when a binary constellation is used for information transmission over a phase noise channel
\cite{Lau2007,Beygi2011,Kayhan2012,Olivares2013}. The described strategy is based on a generalization of a receiver proposed by Kennedy \cite{Kennedy1973} in which a displacement of the constellation in the complex amplitude plane is followed by photon number resolved direct detection.

\section{\uppercase{SQL for binary constellations}}
\begin{figure}
	\centering
	\includegraphics[width=.8\linewidth]{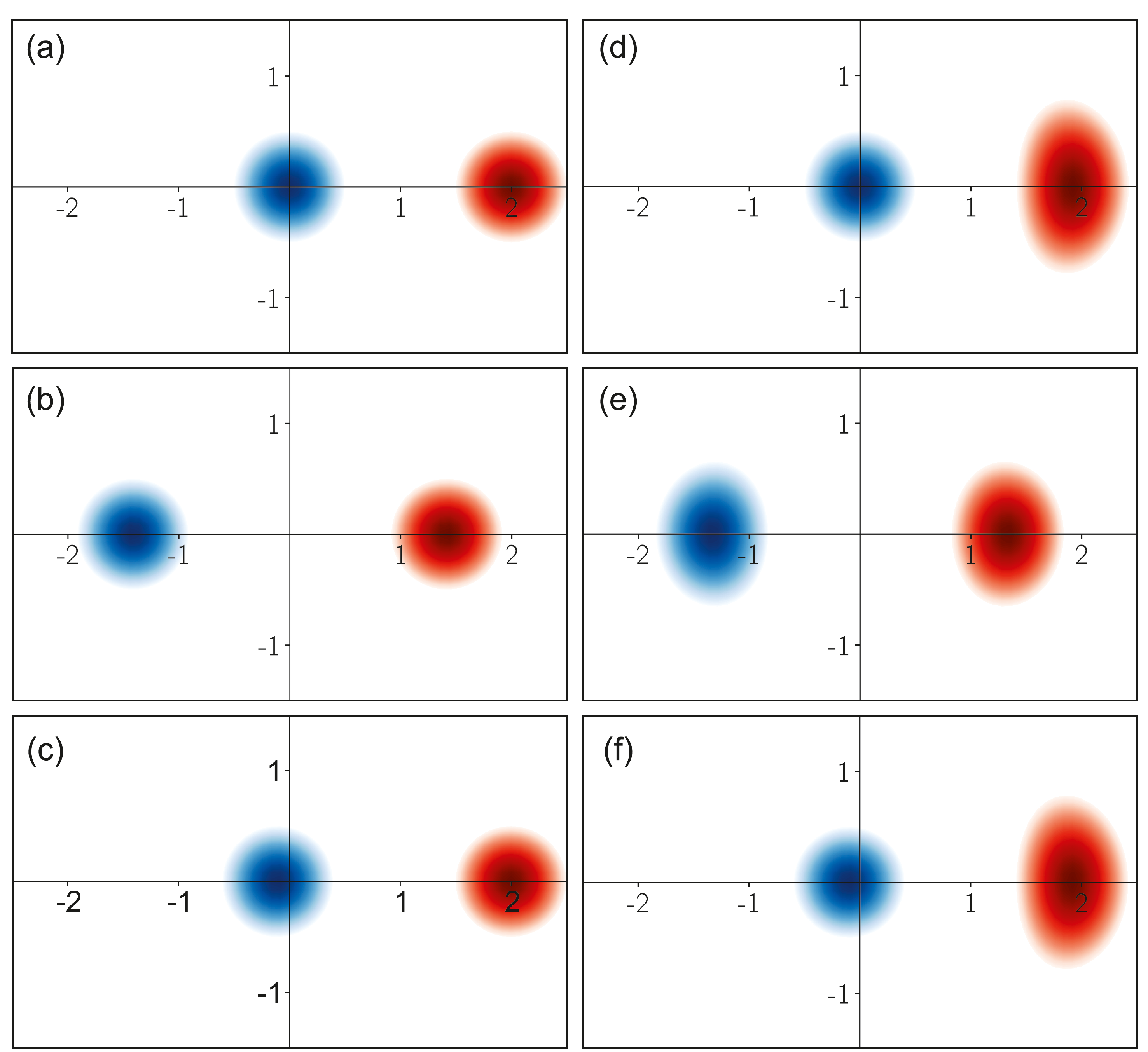}

\bigskip

\caption{(a) On-off keying, (b) binary phase shift keying, and (c) an intermediate constellation for the average photon number $\bar{n}=2$. (d)--(f) The same constellations subjected to Gaussian phase noise with strength $\sigma=0.45$~rad. For this noise value, the intermediate constellation shown in (c),(f) minimizes the error probability when the receiver discussed in section~\ref{Sec:Receiver} is used.}
\label{Fig:Constellation}
\end{figure}

\begin{figure}
	\centering
	\includegraphics[width=.8\linewidth]{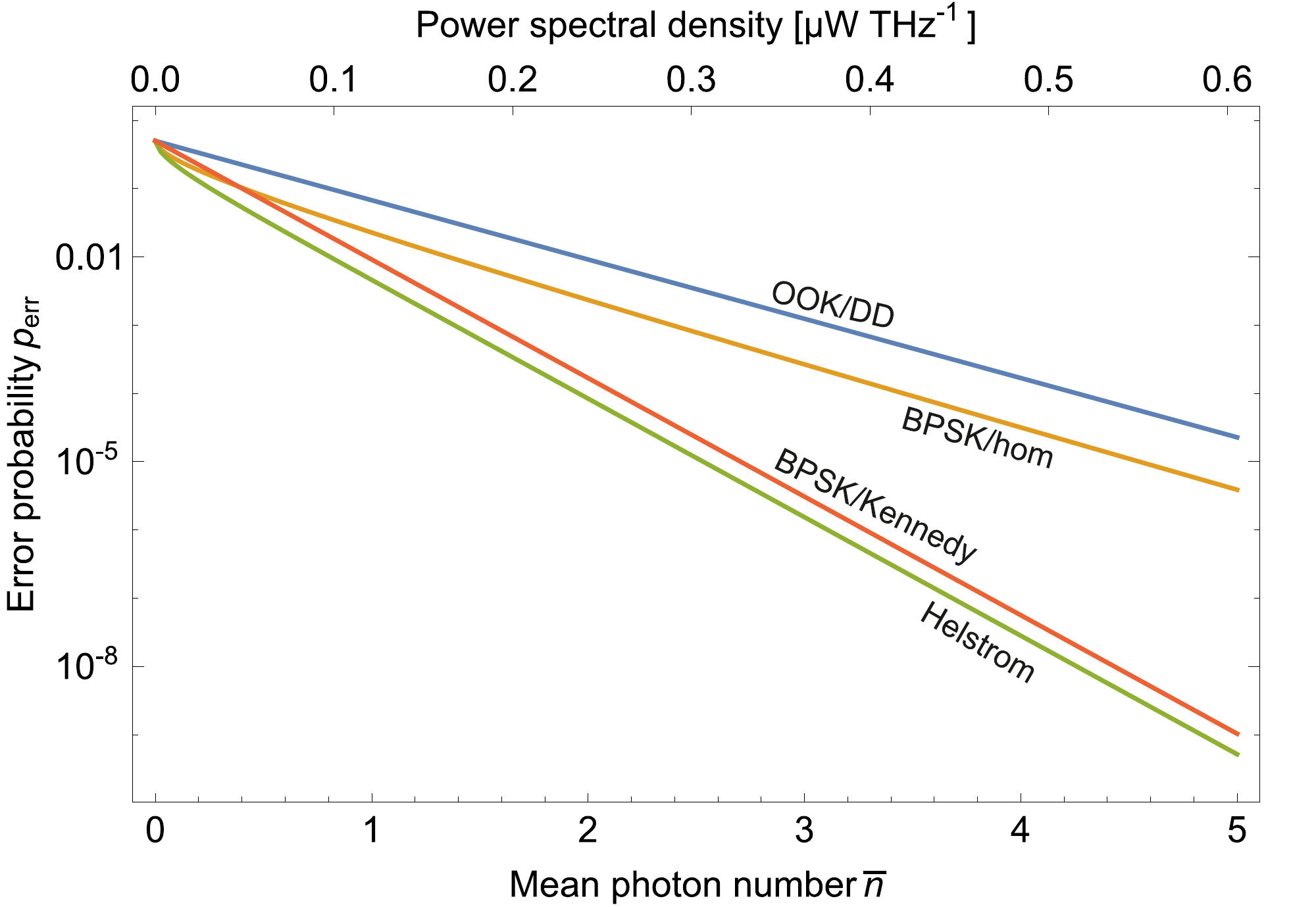}

\bigskip

\caption{The error probability as a function of the average photon number for on-off keying with direct detection (OOK/DD), binary phase shift keying with homodyne detection (BPSK/hom) or the Kennedy receiver (BPSK/Kennedy), as well as the Helstrom bound (Helstrom). For reference, the abscissa is calibrated also in the signal power spectral density at the wavelength $1550$~nm.}
\label{Fig:ErrorProbabilities}
\end{figure}

\noindent
A binary constellation is formed by a pair of equiprobable complex field amplitudes $\alpha_0$ and $\alpha_1$ representing the two bit values. The amplitudes will be specified here in units such that $|\alpha_0|^2$ and $|\alpha_1|^2$ give the mean photon number carried by each symbol. The constraint on the average optical power of the constellation reads
\begin{equation}
	\bar n = \frac{1}{2} \left(\abs{\alpha_0}^2 + \abs{\alpha_1}^2\right).
\label{Eq:PowerConstraint}
\end{equation}
The mean photon number per symbol $\bar{n}$ is a function of the signal power spectral density. A generic example is the on-off keying (OOK) constellation with $\alpha_0=0$ and $\alpha_1 = \sqrt{2\bar{n}}$ shown in Fig.~\ref{Fig:Constellation}(a). The conventional method to read out the bit value carried by an OOK symbol is to measure the signal intensity using direct detection (DD). At the fundamental level, the outcome of the standard measurement of light intensity based on the photoelectric effect has a granular form of individual photocounts. In an idealized scenario when no technical noise is present, the inferred bit value depends on whether at least one photocount has been generated over the symbol time frame or none at all. Errors arise from the fact that detection of a coherent pulse with an amplitude $\alpha_1$ is described by a Poisson point process with the overall zero-count probability equal to $\exp(-|\alpha_1|^2)$. Because errors occur only for the $\alpha_1$ symbol, the average error probability for the OOK/DD scheme reads
\begin{equation}
p_{\text{err}}^{\text{OOK/DD}} = \frac{1}{2}\exp(-|\alpha_1|^2) = \frac{1}{2}\exp(-2\bar{n}).
\label{Eq:perrOOKDD}
\end{equation}
This expression, plotted in Fig.~\ref{Fig:ErrorProbabilities}, defines the SQL for the OOK constellation. Idealized photodetection with unit quantum efficiency has been assumed.
Non-unit efficiency can be taken into account in a straightforward manner by rescaling input complex amplitudes.

Fig.~\ref{Fig:Constellation}(b) depicts the binary phase shift keying (BPSK) constellation with $\alpha_1 = - \alpha_0 = \sqrt{\bar{n}}$. In this case, the bit value can be determined by measuring the in-phase $I$ quadrature. Shot-noise limited homodyne detection of the $I$ quadrature
is characterized by the probability density function
\begin{equation}
p(x|\alpha) = \frac{1}{\sqrt{\pi}} \exp[-(x-\sqrt{2} \text{Re}\alpha)^2].
\end{equation}
The quadrature value $x$ is specified in dimensionless units such that the shot noise variance is $(\Delta x)^2 = 1/2$. The SQL error probability for the BPSK constellation readout using shot-noise limited homodyne detection is given by $\int_{-\infty}^{0} dx \, p(x|\sqrt{\bar{n}}) = [1-\textrm{erf}(\sqrt{2\bar{n}})]/2$, where erf is the error function. This is lower compared to the OOK/DD combination for the same average optical power $\bar{n}$,  as seen in Fig.~\ref{Fig:ErrorProbabilities}.

\section{\uppercase{Helstrom limit}}
\noindent
From a general perspective, one can consider a receiver directly producing a binary outcome corresponding to the two constellation symbols. Ultimately, the error probability attainable with such a receiver is limited by quantum fluctuations inherently present in the electromagnetic field rather than noise specific to any concrete detection scheme. To incorporate quantum fluctuations in the analysis, one needs to resort to the concept of coherent states, which provide a complete quantum mechanical representation of an electromagnetic field waveform with a well defined complex amplitude \cite{Glauber1963}. Interestingly, minimization of the error probability over all detection strategies permitted in quantum mechanics yields the result in a closed analytical form. The expression is the celebrated {\em Helstrom bound} \cite{Helstrom1976}, which for a pair of coherent states
with amplitudes $\alpha_0$ and $\alpha_1$ takes the form
\begin{equation}
p_{\text{err}} \ge [1-\sqrt{1-\exp(-|\alpha_1 - \alpha_0|^2)}]/2.
\label{Eq:Helstromnoiseless}
\end{equation}
Note that Planck's constant is implicitly present in this formula owing to the choice of units for the complex field amplitudes $\alpha_0$ and $\alpha_1$. Because the Helstrom bound depends only on the distance $|\alpha_1-\alpha_0|$ between the two amplitudes in the complex plane, under the average power constraint defined in Eq.~(\ref{Eq:PowerConstraint}) the minimum value is reached for the BPSK constellation for which $|\alpha_1 - \alpha_0|^2 = 4\bar{n}$. When $\bar{n} \gg 1$, the Helstrom bound can be approximated by $[1-\sqrt{1-\exp(-4\bar{n})}]/2 \approx \exp(-4\bar{n})/4$ which yields a more favorable scaling with the signal optical power seen in Fig.~\ref{Fig:ErrorProbabilities}.

\begin{figure}
	\centering
	\includegraphics[width=1.2\linewidth]{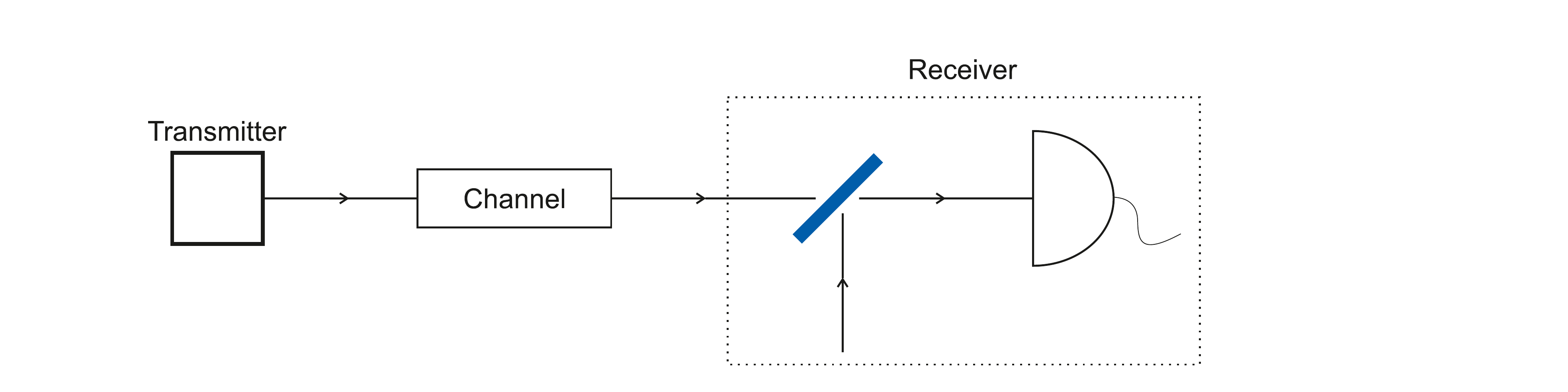}

\caption{A schematic  of the Kennedy receiver. The incoming symbol is displaced by a complex amplitude $\beta$ before measured by a photon number resolving detector. In the original receiver proposed by Kennedy \cite{Kennedy1973} a Geiger-mode detector without photon number resolution was used.}
\label{Fig:Kennedy}
\end{figure}

As pointed out by Kennedy \cite{Kennedy1973}, a nearly optimal discrimination between BPSK symbols can be achieved by displacing the constellation in the complex amplitude plane such that the amplitude of one the symbols is nulled, followed by direct detection. This so-called Kennedy receiver is shown schematically in Fig.~\ref{Fig:Kennedy}. The conditional probabilities for detection outcomes are equivalent to those for the OOK/DD scheme, but with the effective optical power twice as large. Consequently the resulting error probability is $\exp(-4\bar{n})/2$ which differs from the Helstrom bound only by a factor $1/2$. The performance of the Kennedy receiver can be improved for small $\bar{n}$ by fine-tuning the displacement in the complex amplitude plane \cite{Sasaki1996,Takeoka2008,Wittmann2008}. The Helstrom bound is saturated for any average optical power $\bar{n}$ by the Dolinar receiver \cite{Dolinar1973,Cook2007} which uses an adaptive photocounting measurement strategy. The adaptive approach can be extended to larger constellations, such as quadriphase shift keying (QPSK) \cite{Becerra2013}.

\section{\uppercase{Phase noise}}
\noindent
 The effect of a phase noise channel is a multiplication of the complex amplitude $\alpha$ by a random phase factor, $\alpha \rightarrow \alpha e^{i\phi}$. The phase shift $\phi$ will be taken as a Gaussian random variable with variance $\sigma^2$. Consequently, in the presence of phase noise probabilities of detection events need to be averaged according to
\begin{equation}
\bigl\langle \ldots \bigr\rangle_\phi = \frac{1}{\sqrt{2\pi\sigma^2}}  \int_{-\infty}^{\infty} d\phi \, \exp[-\phi^2/(2\sigma^2)] \ldots .
\end{equation}
Obviously, the OOK/DD scheme is not affected by the phase noise and the SQL given in Eq.~(\ref{Eq:perrOOKDD}) holds also in this scenario. The error probability for the BPSK constellation with homodyne detection reads
\begin{equation}
p_{\text{err}}^{\text{BPSK/hom}} = \int_{-\infty}^{0} dx \, \bigl\langle p(x|\sqrt{\bar{n}}e^{i\phi}) \bigr\rangle_\phi.
\end{equation}
Because weak phase noise modifies primarily the $Q$ quadrature, its effect on the error probability in the BPSK/hom scenario is rather minor for $\sigma \ll 1$~rad \cite{Olivares2013}. However, with increasing phase noise strength discrimination of BPSK symbols based on the outcome of homodyne detection becomes more and more difficult. Above a certain threshold it is more beneficial to use the OOK/DD combination. These two alternative schemes are optimal in the respective regions of the phase noise strength also if general binary constellations with the average power constraint given by Eq.~(\ref{Eq:PowerConstraint}) are taken into account, but the readout is restricted exclusively to either direct detection or homodyne detection.

\section{\uppercase{Generalized Kennedy receiver}}\label{Sec:Receiver}
\noindent
Reduction of the error probability below the SQL can be achieved using a generalized version of the Kennedy receiver shown in Fig.~\ref{Fig:Kennedy}. The first ingredient is optimization of the displacement in the complex amplitude plane beyond nulling one of the symbols. The second one is the use of a detector with photon number resolving (PNR) capability. The photocount statistics for a complex amplitude $\alpha$ subjected to phase noise and displaced by $\beta$ reads
\begin{equation}
p_k(\alpha) = \frac{1}{k!} \left\langle  |\alpha e^{i\phi} + \beta|^{2k} \exp (- |\alpha e^{i\phi} + \beta|^{2})
\right\rangle_\phi.
\end{equation}
In the case of a binary constellation formed by amplitudes $\alpha_0$ and $\alpha_1$ the bit value is decided based on whether the photocount number $k$ exceeds a threshold value $K$ or not. Consequently, the error probability is given by the expression:
\begin{equation}
p_{\text{err}} = \frac{1}{2}\sum_{k=0}^{K} p_k (\alpha_1) + \frac{1}{2} \sum_{k=K+1}^{\infty} p_k(\alpha_0).
\end{equation}
The above expression is to be optimized numerically with respect to $\alpha_0$, $\alpha_1$, $\beta$, and $K$ under two constraints. The first one,  given in Eq.~(\ref{Eq:PowerConstraint}), defines the average signal power. The second one is that the integer $K$ has to remain below the maximum number of photons that can be resolved by the photodetector for a single symbol. This PNR ceiling is motivated by limitations of currently available single photon detectors \cite{Mirin2012}.

\section{\uppercase{Results}}
\begin{figure}
	\centering
	\includegraphics[width=.6\linewidth]{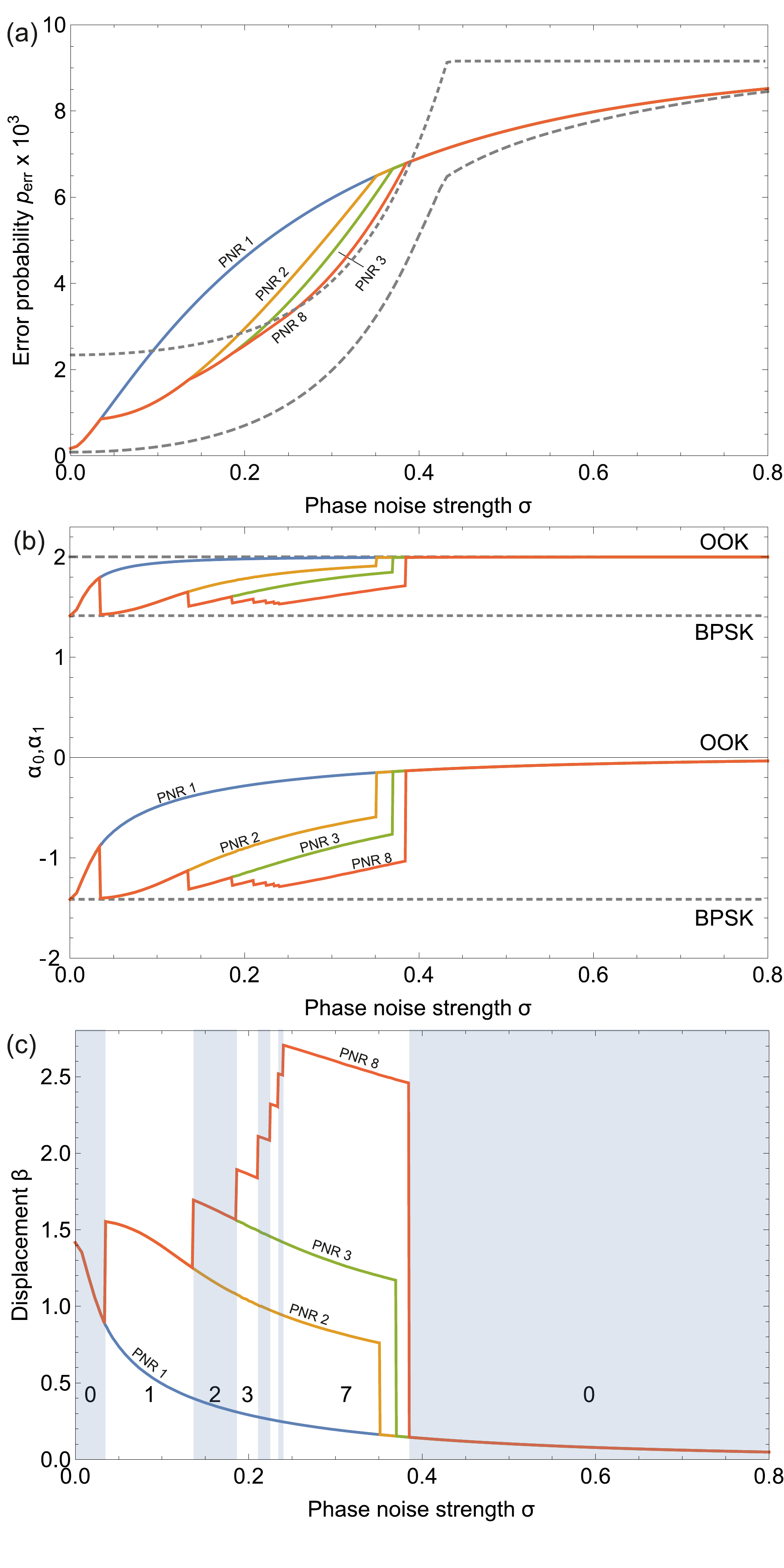}
\caption{(a) Optimized error probability as a function of the noise strength $\sigma$ for the average photon number $\bar{n}=2$ and the photon number resolving (PNR) capability truncated at $1, 2, 3$, and $8$. The dotted and dashed lines depict respectively the standard quantum limit and the Helstrom bound calculated for phase-diffused coherent states. (b) Optimal amplitudes $\alpha_0$ and $\alpha_1$ of the constellation symbols. (c) The displacement $\beta$ of the constellation implemented in the receiver. The alternating white and light gray stripes in the background indicate the optimal threshold photocount number $K$ for the PNR capability truncated at $8$.}
\label{Fig:Optimized}
\end{figure}

\noindent
The results of optimization for $\bar{n}=2$ are shown in Fig.~\ref{Fig:Optimized} as a function of the phase noise strength $\sigma$. It is seen that up to the value $\sigma \lesssim 0.25$~rad it is possible to beat the standard quantum limit, but this requires increasing PNR capability. Above this value the performance becomes comparable to homodyne detection. Note that in this regime the displacement $\beta$ becomes relatively large and the receiver can be viewed as a regular homodyne setup with a strong local oscillator. Interestingly, for $\sigma \gtrsim 0.4$~rad sub-SQL operation can be restored again. In this region, the optimal conventional strategy is the OOK/DD combination. As seen in Fig.~\ref{Fig:Optimized}(b,c), a minor departure from the OOK constellation combined with a small displacement $\beta$ facilitates reduction of the error rate below the SQL. The optimal threshold value for the photocount number is $K=0$, i.e.\ PNR capability is not required. Noteworthily, this strategy approaches the Helstrom bound calculated for coherent states subjected to phase noise. The optimal constellation for $\sigma = 0.45$~rad has been depicted in Fig.~\ref{Fig:Constellation}(c),(f).

\section{\uppercase{Conclusions}}
\noindent
Unconventional detection strategies hold the promise of reducing error rates in optical communication systems below the standard quantum limit. The example presented in this paper shows that such reduction is possible also in the presence of phase noise. The generalized Kennedy receiver can be viewed as an interpolation between direct detection (when the displacement of the complex amplitude is zero) and homodyne detection (realized for a very large displacement). When combined with the optimization of the input binary constellation under the average power constraint, such a receiver exhibits enhanced performance for both low and high noise strengths, in the latter case approaching the quantum mechanical Helstrom bound.

\section*{ACKNOWLEDGEMENTS}
\noindent
We acknowledge insightful conversations with Christoph Marquardt and Matteo G. A. Paris as well as the support of the FNP TEAM project ``Quantum Optical Communication Systems'' and National Science Foundation (NSF) grants (PHY-1653670, PHY-1521016).

\thispagestyle{empty}

\thispagestyle{empty}
% that's all folks
\end{document}